\documentclass[a4paper,twocolumn,aps,prd,nolongbibliography,superscriptaddress,showpacs,showkeys,amsmath,amssymb,floatfix,nofootinbib]{revtex4-1}
\usepackage{lmodern}
\usepackage{yfonts}
\usepackage{graphicx}

\usepackage[T1]{fontenc}
\usepackage[utf8]{inputenc}
\usepackage[usenames]{xcolor}
\usepackage[unicode=true,pdfusetitle,
bookmarks=true,bookmarksnumbered=true,bookmarksopen=true,bookmarksopenlevel=1,
 breaklinks=false,pdfborder={0 0 0},backref=false,colorlinks=true]
 {hyperref}
\hypersetup{
 citecolor=blue,filecolor=blue,linkcolor=black,urlcolor=blue}
\usepackage{ulem}
\makeatletter

\pdfpageheight\paperheight
\pdfpagewidth\paperwidth

 \@ifundefined{textcolor}{}
 {%
   \definecolor{BLACK}{gray}{0}
   \definecolor{WHITE}{gray}{1}
   \definecolor{RED}{rgb}{1,0,0}
   \definecolor{green}{rgb}{0,0.7,0}
   \definecolor{BLUE}{rgb}{0,0,1}
   \definecolor{CYAN}{cmyk}{1,0,0,0}
   \definecolor{MAGENTA}{cmyk}{0,1,0,0}
   \definecolor{YELLOW}{cmyk}{0,0,1,0}
      \definecolor{myred}{rgb}{0.7,0,0.1}
 }
 
\renewcommand{\[}{\begin{equation}}
\renewcommand{\]}{\end{equation}} 
\usepackage{array}
\makeatother

\newcommand{\be}{\begin{equation}}
\newcommand{\ee}{\end{equation}}
\newcommand{\beqa}{\begin{eqnarray}}
\newcommand{\eeqa}{\end{eqnarray}}
\newcommand{\bn}{{\mathbf n}}
\newcommand{\bk}{{\mathbf k}}
\newcommand{\bx}{{\mathbf x}}
\newcommand{\by}{{\mathbf y}}
\newcommand{\ndotk}{(\hat n \cdot \hat k )}
\newcommand{\PP}{\mathcal{P}}
\newcommand{\HH}{\mathcal{H}}
\newcommand{\di}{\text{d}}

\newcommand{\ipartial}{{\mkern4.5mu\mathchar'26\mkern-13.5mu\partial}}
\newcommand{\corr}[1]{\left\langle #1 \right\rangle}
\newcommand{\Hcal}{\mathcal{H}}
\newcommand{\ds}{\mathrm{d}}
\newcommand{\gs}{\mathrm{g}}
\newcommand{\ms}{\mathrm{m}}
\newcommand{\rb}{\boldsymbol{r}}
\newcommand{\nb}{\boldsymbol{n}}
\newcommand{\eb}{\boldsymbol{e}}
\newcommand{\kb}{\boldsymbol{k}}
\newcommand{\vb}{\boldsymbol{v}}

\newcommand{\xb}{\boldsymbol{x}}
\newcommand{\omb}{\boldsymbol{\omega}}
\newcommand{\nablab}{\boldsymbol{\nabla}}
\newcommand{\kat}{\hat{k}}
\newcommand{\de}{\delta}

\newcommand{\pd}{\partial}

\newcommand{\la}{\lambda}
\newcommand{\ep}{\epsilon}
\newcommand{\Si}{\Sigma}
\newcommand{\om}{\omega}
\newcommand{\Om}{\Omega}
\newcommand{\nn}{\nonumber}

\newcommand{\tj}[6]{ \begin{pmatrix}
   #1 & #2 & #3 \\
   #4 & #5 & #6 
\end{pmatrix}}

\newcommand{\ninej}[6]{ \begin{Bmatrix}
   #1 & #2 & #3 \\
   #4 & #5 & #6 
\end{Bmatrix}}

\def\apjl{ApJL }

\def\apjs{ApJS }

\def\aap{A\&A }

\begin{document}

\title{Redshift-space distortions from vector perturbations II: Anisotropic signal}

\author{Vittorio Tansella}
\affiliation{D\'epartment de Physique Th\'eorique and Center for Astroparticle Physics,
Universit\'e de Gen\`eve, Quai E. Ansermet 24, CH-1211 Gen\`eve 4, Switzerland}
\author{Camille Bonvin}
\affiliation{D\'epartment de Physique Th\'eorique and Center for Astroparticle Physics,
Universit\'e de Gen\`eve, Quai E. Ansermet 24, CH-1211 Gen\`eve 4, Switzerland}
\author{Giulia Cusin}
\affiliation{Astrophysics Department, University of Oxford, DWB, Keble Road, Oxford OX1 3RH, UK}
\author{Ruth Durrer}
\affiliation{D\'epartment de Physique Th\'eorique and Center for Astroparticle Physics,
Universit\'e de Gen\`eve, Quai E. Ansermet 24, CH-1211 Gen\`eve 4, Switzerland}
\author{Martin Kunz}
\affiliation{D\'epartment de Physique Th\'eorique and Center for Astroparticle Physics,
Universit\'e de Gen\`eve, Quai E. Ansermet 24, CH-1211 Gen\`eve 4, Switzerland}

\author{Ignacy Sawicki}

\affiliation{CEICO, Institute of Physics of the CAS, Na Slovance 2, 182 21 Praha 8, Czechia}

\begin{abstract}
We study the impact on the galaxy correlation function of the presence of a vector component in the tracers' peculiar velocities, in the case in which statistical isotropy is violated. We present a general framework -- based on the bipolar spherical harmonics expansion --  to study this effect in a model independent way, without any hypothesis on the origin or the properties of these vector modes. We construct six new observables, that can be directly measured in galaxy catalogs in addition to the standard monopole, quadrupole and hexadecapole, and we show that they completely describe any deviations from isotropy. We then perform a Fisher analysis in order to quantify the constraining power of future galaxy surveys. As an example, we show that the SKA2 would be able to detect anisotropic rotational velocities with amplitudes as low as 1\% of that of the vorticity generated during shell-crossing in standard dark matter scenarios.

\end{abstract}
\maketitle

\tableofcontents

\section{Introduction}
\label{sec:intro}
Mechanisms such as topological defects~\cite{Durrer:2001cg,Daverio:2015nva,Lizarraga:2016onn}, magnetic fields~\cite{Durrer:1998ya}, inflation with vector fields~\cite{Ford:1989me,Golovnev:2008cf}, or vector-field-based models of modified gravity~\cite{Jacobson:2000xp,Zlosnik:2006zu,Heisenberg:2014rta,Tasinato:2014eka}, but also the shell-crossing present in concordance cosmology, can generate vector perturbations throughout the history of the Universe and on a wide range of scales.  It is important to properly characterize the signature of these vector degrees of freedom on the  observables of large scale galaxy surveys. The reason for this is twofold. On the one hand, the presence of such vector perturbations -- if not properly taken into account -- will `pollute' (i.e.\ bias) the measurement of the scalar degrees of freedom and act as a source of systematic error. 
On the other hand vector degrees of freedom can leave their imprint on observables which, in turn, can be used to constrain their properties and to study the mechanism that generated them.

Various approaches exist in the literature with the aim of constraining vector-type deviations of the metric and they have mostly focused on the Cosmic Microwave Background (CMB). They can be grouped into three categories: (i) introducing dynamical vector degrees of freedom in the early universe while maintaining isotropy and homogeneity at the background level. Then, one can either maintain statistical isotropy and homogeneity of the perturbations or allow for statistically anisotropic perturbations~\cite{Lim:2004js,Nakashima:2011fu}.
Alternatively (ii), one can deform the isotropy of the cosmological background and therefore constrain its anisotropy, while keeping the matter content standard, making sure that this anisotropy decays with time \cite{Saadeh:2016bmp}. Finally (iii), one can introduce an anisotropy directly in the primordial power spectrum (through some interactions in the early universe, e.g.\ \cite{Ackerman:2007nb,Bartolo:2017sbu}). One then tries to look for `anomalies' in the CMB, such as in, for example,~\cite{Ade:2015hxq}. Signatures of this primordial signal in galaxy surveys have been analyzed in \cite{2010JCAP...05..027P,Jeong:2012df,Shiraishi:2016wec,Sugiyama:2017ggb}.

Additionally, late time non-linear evolution, as simulated in N-body codes, is found to generate vector perturbations of both the metric~\cite{Adamek:2015eda,Adamek:2016zes} and the fluid vorticity~\cite{Pueblas:2008uv,Jelic-Cizmek:2018gdp}. It is interesting to develop statistical tools to measure these vector modes, which are present also in standard $\Lambda$CDM cosmology, and to distinguish them, e.g. from an intrinsic, global anisotropy.

In \cite{Bonvin:2017req} some of us considered the impact of statistically isotropic vector modes in the peculiar velocity field of  galaxies and in particular on the redshift-space distortion (RSD) observed in galaxy surveys. We have found that vector contributions to RSD enter in the monopole, quadrupole and hexadecapole of the galaxy correlation function. While the impact of vector perturbations from topological defects is very small, those from non-linear clustering affect especially the hexadecapole quite strongly, contributing up to 20\% of the total signal on scales smaller than 5$h^{-1}$Mpc. This additional contribution should in principle be detectable with next generation surveys, such as Euclid or the SKA.

In this paper we consider a vector component of the peculiar velocities, which violates statistical isotropy, and study its impact on the galaxy correlation function. This is a natural generalization of the study in \cite{Bonvin:2017req}. We present a general framework suitable specifically to study this effect, with no assumptions on the origin or properties of these vector modes. We show that the anisotropic signal can be completely characterised by six new observables, that can be directly extracted from galaxy catalogs. General results regarding the Fisher analysis of these types of models are also discussed. We investigate the detectability of these contributions, for a specific example, with planned or futuristic galaxy surveys.

This paper can be considered as a contribution to testing the cosmological principle. In particular we want to develop tests of statistical isotropy using large-scale structure (LSS) observations. While it is clear that our Universe is not strongly anisotropic, a small anisotropy is still compatible with, if not favoured by the analysis of CMB anisotropies and polarisation~\cite{Ade:2015hxq}. This might be due to e.g. a small global magnetic field or some slight anisotropy which remained after inflation. In this work we do not make assumptions on the model responsible for the global anisotropy in the vector sector but we want to investigate its observational consequences.  We study the situation where scalar perturbations are still statistically isotropic but vector perturbations are not. It will be interesting not only to study whether LSS also favours a slight anisotropy of the Universe but whether the characteristics of any such anisotropy are in agreement with the one of the CMB. Furthermore, LSS observations allow for a tomographic approach, i.e.\ we can observe many different redshifts, making it easier to overcome limitations from cosmic variance.

The paper is structured as follows: In section \ref{sec:vec_gen} we detail the general anisotropic structure of vector perturbations. In \ref{seccf} we study the effects of a vector component in the velocity field on the two-point function and present a suitable decomposition to describe it. Finally, in section \ref{sec:fisher}, we forecast the constraints on the anisotropic parameters for upcoming clustering surveys.

\section{Vector Contribution to Galaxy Velocities}
\label{sec:vec_gen}

In this work we assume that our Universe shows signs of a violation of statistical isotropy, manifesting itself by the presence of vector modes in the peculiar velocity of tracers. We investigate how galaxy catalogs can be used, independently from other probes, to constrain the amplitude of these anisotropies. We therefore model  our Universe as a perturbed Friedman-Lema\^\i tre universe, with a metric given by
\begin{align}\label{eq:metric}
\mathrm{d}s^2=a^2\Big[&-(1+2\Psi)\mathrm{d}\tau^2 - \Sigma_i \mathrm{d}\tau\mathrm{d}x^i \\\nonumber& + (1-2\Phi)\mathrm{d}x_i\mathrm{d}x^i\Big]\,.
\end{align}
Here $\Phi$ and $\Psi$ are the standard Newtonian-gauge scalar potentials, and $\Sigma_i$ is a pure vector fluctuation, $\partial_i\Sigma^i=0$, related to frame dragging\footnote{We have fixed the gauge such that the $0i$ component of the metric has no scalar contribution and  the vector part of the $ij$ component vanishes. We also neglect gravitational waves (tensor perturbations).}. %
We define $\mathcal{H} =\dot a/a =a H $ to be the conformal Hubble parameter.

The general velocity field for galaxies located at position $\boldsymbol{r}$ at conformal time $\tau$, $v^i(\boldsymbol{r},\tau)$, can be decomposed into a scalar (potential) part, $v$, and a pure vector part, $\Omega^i$, with $\partial_i\Omega^i =0$, 
\begin{equation}
v^{i}\equiv \partial^{i}v+\Omega^{i} \,.\label{eq:vel}
\end{equation}
The gauge-invariant relativistic vorticity~\cite{Rbook} can be obtained by lowering the index of $\Omega^i$ with the perturbed metric. The relativistic vorticity is often denoted $\Omega_i$ (e.g.\ in~\cite{Lu:2008ju,Rbook}) and it is an additional rotational velocity over and above the frame-dragging effect. In this paper we denote it by $\tilde\Om_i \equiv g_{ij}\Om^j/a =  a\de_{ij}(\Om^j-\Si^j)$ for clarity\footnote{The difference between $a\Om^j$ and $\tilde \Om_j$ is only relevant on large scales.}. We mainly concentrate on $\Omega^i$ as it is the velocity with an upper index that is relevant for us and we use the notation $\Om_i = \de_{ij}\Om^j \equiv \Om^i$.

We assume that galaxies move on time-like geodesics of the metric, i.e.\ they obey the Euler equation.
Then, to first order in perturbation theory we can write, for perfect fluids,
\begin{equation}
\dot \Omega_i - \dot \Sigma_i  + \mathcal{H} (\Omega_i-\Sigma_i) = 0 \,,\label{eq:vecgeo}
\end{equation}
which is equivalent to $\partial_\tau {\tilde\Om_i} =0$. Hence vorticity is conserved. This is not only true within linear perturbation theory but also in full General Relativity as long as matter can be described as a perfect fluid~\cite{Lu:2008ju}.
The $0i$ component of the energy momentum tensor of a perfect fluid is given by 
\be
 T^{\,i}_{0\,(V)}=[(\rho+P)v^i]_{(V)} = T^{\,i}_{0}-T^{\,i}_{0\,(S)} \,.
\ee
Taking the curl of this equation the scalar part vanishes and we obtain
\beqa
\ep_{ijk} (T^{\,j}_{0})_{,k} &=&\ep_{ijk} [(\rho+P)v^j]_{,k} \nonumber \\ 
&=& \left(\vb\wedge\nablab(\rho+P)\right)_i + (\rho+P)(\nablab\wedge\vb)_i\,. \label{e:curlT0i}
\eeqa
Only the vector velocity $\Om^j$ contributes to the second term, while the first term is non-vanishing when the gradient of the density fluctuations in not parallel to the velocity. This also happens in perfect fluids at second order in perturbation theory, see e.g.~\cite{Adamek:2016zes}. At second order therefore, despite vorticity conservation,  vector perturbations of the metric are generated, which induce effects like frame dragging. It has been shown recently~\cite{Jelic-Cizmek:2018gdp} that the vector potential found in relativistic numerical simulations is actually mainly due to the first term of \eqref{e:curlT0i} and not to vorticity which is also induced in N-body simulations.

The perfect fluid description is just an approximation when we want to describe the motion of dark matter (or galaxies). In the real Universe, dark matter particles are free-streaming, i.e.\ they move on geodesics. As soon as shell crossing occurs, velocity dispersion can no longer be neglected and vorticity is generated for the fluid of the averaged dark matter particles (or galaxies). In~\cite{Cusin:2016zvu}, the vorticity generation from large-scale structure was modelled by including velocity dispersion using a perturbative approach.

Clearly, even if in the standard $\Lambda$CDM model vector perturbations are generated by non-linearities, they are statistically isotropic. In this work we assume that the vectorial part of the peculiar velocity in Eq.~\eqref{eq:vel} acquires an anisotropic component.

\subsection{Tensor structure of vector perturbations}

We summarise here the discussion we presented in section 2.2 of \cite{Bonvin:2017req}. 

In order to compute the two-point correlation function of galaxies (2PCF), we need a model for the two-point auto-correlation of the vector velocity, $\corr{\Omega_i\Omega_j}$ and its cross-correlation with the dark matter overdensity $\corr{\delta_\ms\Omega_i}$. We will characterise their structure in Fourier space, with our Fourier transform convention fixed by
\[
f(\kb) = \int d^3 r f(\rb) e^{-i \kb . \rb} \, .
\]

\begin{enumerate}
\item The auto-correlation of the vector field takes the general form
\begin{align}
&\langle\Omega_{i}(\kb)\Omega_{j}(\kb')\big\rangle= (2\pi)^{3}\delta^{(3)}(\kb+\kb')\nn\\
&\qquad \times 
\left[W_{ij}(\bk)P_{\Omega}(k) +i\alpha_{ij}(\bk)P_A(k)\right] \,,
\label{omom}
\end{align}
where $P_{\Omega}(k)$  and $P_A(k)$ contain information about the amplitude of the vector field. The Dirac delta function appearing in the above equation, $\delta^{(3)}(\kb+\kb')$, is a consequence of statistical homogeneity, and if we assume that scalar spectra are isotropic, the amplitudes,  $P_{\Omega}(k)$  and $P_A(k)$, depend on $\bk$ only through its absolute value $k\equiv |\mathbf{k}|$.  One might think it would be more natural for an anisotropic spectrum to show an anisotropy also in $P_{\Omega}(\kb)$. However, in a real observation, the power spectrum is usually obtained by averaging the squared Fourier modes over directions. Here we mimic this by considering $P_\Om$ and $P_A$ to be functions of the modulus $k$ only. In practice, these are the direction averaged spectra.
For scalar perturbations, this averaging removes all signs of an anisotropy, for vector perturbations this is, interestingly, not the case as we show in this paper.

The tensors $W_{ij}$ and $\alpha_{ij}$ are, respectively, symmetric and anti-symmetric tensors, that encode the dependence on direction.
Since $\Omega_i$ is a pure vector field, $W_{ij}$ and $\alpha_{ij}$ must satisfy $k^i W_{ij}=k^j W_{ij}=k^i\alpha_{ij}=k^j \alpha_{ij}=0$. 
The $P_A$-term is parity odd while the $P_\Omega$-term is parity even. If no parity violating processes occur in the Universe we may set $P_A=0$. The  tensorial form for $\alpha_{ij}$ is completely fixed by anti-symmetry and transversality, 
\begin{equation}\label{eq:aij}
\hspace{0.3cm}\alpha_{ij}= \alpha\varepsilon_{ijm}\hat k^m \,.
\end{equation}
The most general form for $W_{ij}$ is then
\begin{align}
\hspace{0.7cm}W_{ij}=~&\frac{\omega}{2}\left(\delta_{ij}-\hat{k}_{i}\hat{k}_{j}\right)+ \omega_{ij}^A\,,\label{eq:wij}
\end{align}
where we have decomposed the tensor into its trace $\omega$ and trace-free part 
\begin{align}\label{omegaA}
\omega^A_{ij}&=\omega_{ij}-\omega_{il}\hat{k}^{l}\hat{k}_{j}-\omega_{lj}\hat{k}^{l}\hat{k}_{i} \nonumber \\ 
& \quad +\omega_{lm}\hat{k}^{l}\hat{k}^{m}\hat{k}_{i}\hat{k}_{j} \,,
\end{align}
with $\omega^i_i=0$.  As usual $\hat{\kb}$ denotes the unit vector in the direction of the vector $\kb$. The first term of~\eqref{eq:wij} respects statistical homogeneity and isotropy, whereas the second one is non-zero only when isotropy is violated. In what follows, we  absorb the trace $\omega$ into the normalisation of the power spectrum $P_\Omega$ in Eq.~(\ref{omom}).  Note that in general the isotropic and anisotropic contribution do not need to have the same amplitude $P_\Omega(k)$: in this sense one can use $\omega(k)$ to parametrise the difference between  $P_\Omega^\text{(iso)}$ and  $P^\text{(ani)}_\Omega$. Interestingly, the only possible parity odd term given in  \eqref{eq:aij} is statistically isotropic.

The symmetric tensor $\omega_{ij}$ can be diagonalised or, equivalently, decomposed into a sum of the tensor products of its orthonormal eigenvectors $\hat\omega_i^{I}$, 
\be
\omega_{ij}=\sum_{I=1}^{3}\lambda_{I}\hat\omega_{i}^{I}\hat\omega_{j}^{I}\,,
\label{wdec}
\ee
where the eigenvalues satisfy $\sum_I\lambda^I=0$.

\item The cross-correlation with dark matter can be non-zero only if statistical isotropy is violated. Assuming that the vector field is fluctuating in some fixed direction $\hat{\omb}$, the cross-correlation takes the form
\begin{align}
\hspace{0.5cm}\corr{\delta_\ms(\kb)\Omega_{i}(\kb')}  &= (2\pi)^{3} W_i P_{\delta\Omega}(k)\delta^{(3)}(\kb +\kb') \,, 
\end{align}
where $W_i$ is transverse since $\Omega_i$ is a pure vector field i.e.\ divergence free. A non-vanishing  $\corr{\delta_\ms\Omega_{i}}$ always defines a preferred spatial direction $\hat\omega_i$ and therefore violates statistical isotropy.	
\end{enumerate}

\section{Correlation function}\label{seccf}

Galaxy number counts are observed in redshift-space, rather than in real-space. The leading correction arising from the fact that we observe on the light-cone is the Kaiser term, or redshift-space distortion \cite{Kaiser:1987qv}, which is included in the number counts $\Delta$ as
\[\label{Delta}
\Delta(\rb)=\delta_\gs(\rb)-\frac{1}{\mathcal{H}}n^{i}\partial_{i}(n^{j}v_{j}(\rb))\,.
\]
Here $\delta_g$ is the tracer's density perturbation, related to the dark matter density perturbation via the bias expansion $\delta_g \simeq b\cdot \delta_m + ...$, and  $v_i$ is the peculiar velocity field. We  have also defined the line-of-sight direction $\nb$ as 
\[
\nb\equiv\frac{\rb}{r} \,,
\]
i.e.\ the unit vector in the direction of the galaxy lying at $\boldsymbol{r}$, with the observer
located at $\boldsymbol{r}=0$.
Splitting the velocity into the scalar and vector parts, as in Eq.~\eqref{eq:vel}, we have
\begin{equation}
\Delta(\rb)=\delta_\gs(\rb)-\frac{1}{\mathcal{H}}n^{i}n^{j}\big(\partial_{i}\partial_{j}v(\rb)+\partial_{i}\Omega_{j}(\rb)\big) \, .\label{eq:deltaz}
\end{equation}
The effects of vector perturbations in the general relativistic number counts were derived in \cite{Jeong:2014ufa} and studied in detail in~\cite{Durrer:2016jzq}, where it was found that -- akin to scalar perturbations -- redshift-space distortion is the dominant effect. Since in the relativistic angular power spectra, $C_\ell(z_1,z_2)$, the RSD cannot easily be extracted, we study here the impact of the vector modes on the two-point correlation function of galaxies. In this study we neglect both the sub-dominant vector relativistic corrections from~\cite{Durrer:2016jzq} and the scalar relativistic corrections derived in~\cite{Yoo:2009au,Bonvin:2011bg,Challinor:2011bk,Baldauf:2011bh,Jeong:2011as}.

The two-point correlation function is defined as
\[\label{corr-def}
\xi(\rb_1,\rb_2,z_1,z_2)=\langle \Delta(\rb_1,z_1)\Delta(\rb_2,z_2) \rangle\, .
\]
Without redshift-space distortion, and neglecting subdominant evolution effects, the correlation function depends only on the galaxies' separation
\[
x \equiv |\rb_1 - \rb_2|\, ,
\]
and on the mean distance of the pair from the observer $\bar r = \frac{1}{2} (r_1 + r_2)$ or, equivalently, its mean redshift $\bar z =  \frac{1}{2} (z_1 + z_2)$.
Redshift-space distortion introduces an additional dependence on the orientation of the pair with respect to the line-of-sight $\bn$ (we work in the small angle or flat-sky limit where we neglect the difference between the line-of-sight to $\rb_1$ and $\rb_2$). It is customary to expand $\xi$ in a basis of Legendre polynomials so that, in the flat-sky approximation, $\nb_1=\nb_2=\nb$ we can write
\be\label{corrP}
\xi(\bar z, \bx, \bn)=\sum_{\ell}\xi_{\ell}(\bar z, x) \mathcal{P}_{\ell}(\mu)\,,
\ee
where $\mathcal{P}_\ell$ is the Legendre polynomial of degree $\ell$ and $\mu = \boldsymbol{n}\cdot\hat{\boldsymbol{x}}$, with  $\hat{\boldsymbol{x}}$ being the direction of the vector connecting the two galaxies.

Let us now review the standard flat-sky expression for the correlation function in the presence of scalar perturbations (see e.g. \cite{Tansella:2017rpi} for details). We will use this result both for comparison with the vector case and to compute our covariance matrix in section~\ref{sec:fisher}. Including the Kaiser term we write 
\begin{align}
\xi^{\text{iso}}_\text{(s)}(\bar z, x,\mu)&= c_0(\bar z) C_{0}(\bar z,x) \label{xiscalar}-c_2(\bar z)C_{2}(\bar z,x)\mathcal{P}_{2}(\mu)  \notag \\
& \quad +c_4(\bar z) C_{4}(\bar z,x)\mathcal{P}_{4}(\mu)  \,. \qquad
\end{align}
We can identify the multipole coefficients in  Eq.\ (\ref{corrP}) as 
\be
\xi_{\ell}(x, \bar{z})=i^{\ell} c_\ell(\bar z) C_{\ell}(\bar z,x)\,.
\ee
We have also defined 
\begin{equation}
C_\ell(\bar z,x)=\int\frac{\mathrm{d}k}{2\pi^2}\,k^{2}P(\bar z, k)j_\ell(kx) \,,
\end{equation}
together with the coefficients:
\begin{align}
c_0 &=  b^2+\frac{2}{3}bf +\frac{f^2}{5} \,,\\
c_2 &=  \frac{4}{3} bf +\frac{4}{7} f^2 \,, \\
c_4 &= \frac{8}{35} f^2\,.
\end{align}
Here $j_n$ is the $n$-order spherical Bessel function, $f$ is the growth rate, 
$f\equiv d\ln D_1/d\ln a$ (with $D_1$ being the linear growth function), 
and $P(\bar z, k)$ is the matter power spectrum at redshift $\bar z$. We have made the standard assumption that the galaxy bias $b$ is deterministic and, like the growth rate $f$ in $\Lambda$CDM, it is scale independent.

We now turn to the study of the vector component. We split the vector contribution to the correlation function into a statistically isotropic and anisotropic part
\be\label{corr-split}
\xi_\text{(v)}=\xi_\text{(v)}^{\text{iso}}+\xi^{\text{ani}}_{\text{(v)}}\,,
\ee
where we have emphasized that the source of violation of statical isotropy comes from the vector sector. First, we summarise the structure of the isotropic contributions to the correlation function coming from vector perturbations and we then propose a general framework to compute the anisotropic part. 

The new vector contribution to the correlation function in Eq.~(\ref{corr-split}) comprises three terms:
\begin{enumerate}
\item Cross-correlation with the density
\item Cross-correlation with the scalar velocity
\item Auto-correlation.
\end{enumerate}
The first two vanish in flat sky since they are odd under $\nb\rightarrow -\nb$ and $\xi$ is evidently even, see \cite{Bonvin:2017req}. Hence combining Eqs.~(\ref{eq:deltaz}) and~(\ref{corr-def}) we write
\begin{align}
\xi_{\text{(v)}}(x) = \frac{1}{\Hcal(z_1)\Hcal(z_2)}&\int\!\frac{\ds^{3}k}{(2\pi)^{3}} k^2(\nb_1\cdot\hat\kb)(\nb_2\cdot\hat\kb)\\
&\times n_1^{i}W_{ij}(\hat\kb)n_2^{j}P_{\Omega}(k)e^{i\boldsymbol{k}\cdot\boldsymbol{x}}\, . \nonumber
\end{align}
This object has a complicated tensor structure, which characterises the anisotropy of the vector field. However, when isotropy is assumed we simply write  $W_{ij}=\delta_{ij}-\kat_i\kat_j$, so that, see~\cite{Bonvin:2017req},
\begin{align}
\xi_{\text{(v)}}^{\text{iso}}= \frac{1}{\Hcal^2}\int\!&\frac{\ds^{3}k}{(2\pi)^{3}} k^2(\nb\cdot\hat\kb)^2\big(1+(\nb\cdot\hat\kb)^2\big)\\
&\times P_{\Omega}(k)e^{i\boldsymbol{k}\cdot\boldsymbol{x}}\, . \nonumber
\end{align}
Rewriting the $\nb\cdot\hat\kb$ contributions in terms of Legendre polynomials and integrating over the direction of $\kb$, we obtain for the isotropic contribution~\cite{Bonvin:2017req}
\begin{align}
\xi_{\text{(v)}}^\text{iso}&(\bar z, x, \mu) =\frac{2}{15}\mathcal{P}_{0}(\mu)C_{0}^{\Omega}(x)\label{xivec}\\
&-\frac{2}{21}\mathcal{P}_{2}(\mu)C_{2}^{\Omega}(x) 
-\frac{8}{35}\mathcal{P}_{4}(\mu)C_{4}^{\Omega}(x)\,, \notag
\end{align}
with
\begin{equation}
C_{n}^{\Omega}(x)=\frac{1}{2\pi^{2}}\frac{1}{\Hcal^2}\int\mathrm{d}k\,k^{4}P_{\Omega}(k)j_{n}(kx) \,.\label{COmega}
\end{equation}
Notice here the extra $k^2$ factor multiplying $P_{\Omega}$, which is absorbed in the scalar case when the velocity power spectrum is re-expressed in terms of the density power spectrum.

Statistically isotropic vector perturbations modify the shape of the multipoles coefficients in the Legendre expansion of $\xi$. One can  estimate this effect and study its detectability. This was the strategy followed in~\cite{Bonvin:2017req}. In the anisotropic case however, the standard multipole expansion fails to capture the additional angular dependence encoded in $W_{ij}$. In the next section we therefore consider the decomposition of this dependence into  bipolar spherical harmonics (BipoSH)~\cite{varshalovich1988quantum}.

\subsection{Statistically anisotropic contribution}

When statistical isotropy is violated, the correlation function is no longer only a function of $\mu=\nb\cdot\hat\xb$. Therefore, the standard expansion in Legendre polynomials does not properly describe the angular dependence of $\xi$. The correlation function can however be expanded in terms of the orthonormal set of bipolar spherical harmonics (BipoSH). Since this approach captures an arbitrary angular dependence of the observable under consideration, it has been used in cosmology to analyse CMB~\cite{Hajian:2003qq,Hajian:2005jh,Basak:2006ew,Pullen:2007tu, Cusin:2016kqx, 2012PhRvD..85b3010B} and LSS~\cite{Heavens:1994iq,Hamilton:1995px,Szalay:1997cc,Szapudi:2004gh,Papai:2008bd,Bertacca:2012tp,Shiraishi:2016wec,Bartolo:2017sbu,Sugiyama:2017ggb} data.

In the small angle approximation the correlation function depends on two directions $\xi(\bn, \bx)$, we hence expand
\be
\xi (\bx,\bn, \bar{z}) = \sum_{\ell \ell' \, J M} \, \xi_{\ell\ell'}^{JM}( x, \bar{z} ) X^{JM}_{\ell\ell'} (\hat \bx,\bn) \,,\label{corrB}
\ee
with
\be
\begin{split}
X^{JM}_{\ell\ell'} (\hat \bx,\bn) &= \{Y_\ell (\hat \bx) \otimes Y_{\ell'}(\bn) \}_{JM} \\&= \sum_{mm'} \mathbf{C}_{\ell m \ell'm'}^{JM} Y_{\ell m} (\hat \bx) Y_{\ell'm'}(\bn) \,,
\end{split}
\ee
where $\mathbf{C}_{\ell m \ell'm'}^{JM}$ are the Clebsch Gordan coefficients which 
are related to the Wigner 3j symbols by, see~\cite{AW}, 
\be
\mathbf{C}_{\ell m\ell'm'}^{JM} = (-)^{\ell-\ell'+M} \sqrt{2J+1} \tj{\ell}{\ell'}{J}{m}{m'}{-M}  \,.
\ee
In other words, $X^{JM}_{\ell\ell'} (\hat \bx,\bn)$ isolates the total angular momentum $J$ and helicity $M$ contribution. 
The useful property of the BipoSH $X^{JM}_{\ell\ell'}$ is that they filter the isotropic signal into the $J=0$ mode and any non-zero coefficient with $J >0$ indicates anisotropy. In fact, if there is no anisotropic signal in the power spectrum, i.e. if $\xi$ depends on $\bn$ only via $\mu=\hat\bx\cdot\bn$, we can compute the coefficients via
\be
 \xi_{\ell\ell'}^{JM} = \int d \Omega_\bn \int d \Omega_\bx \,\xi(\bx,\bn,\bar z) X^{JM*}_{\ell\ell'} \,,
\ee
and we simply obtain
\be
\xi_{\ell\ell'}^{JM}(x, \bar{z}) = \frac{4 \pi}{\sqrt{2\ell+1}} \xi_\ell(x, \bar{z}) \delta_{J,0} \delta_{M,0} \delta_{\ell, \ell'}\,,
\ee
recovering the expansion of Eq.\ (\ref{corrP}). In particular, we see that no off-diagonal component is generated (we have $\ell=\ell'$) and that all the isotropic signal is contained in the $J=0$ coefficient. On the other hand if anisotropy is included we will generate $ J \ge 1$ and $\ell\ne\ell'$ modes. Therefore, to search for anisotropy we only look at the  $ J \ge 1$ modes, and we set $\xi = \xi_{\text{(v)}}^\text{ani}$ in the expansion of Eq.~(\ref{corrB}).

We focus on the computation of the statistically anisotropic contribution to the galaxy correlation function (\ref{corr-def}). To this end, it is useful to compute the anisotropic contribution to the power spectrum of (\ref{Delta}) and then Fourier transform it. Explicitly, the Fourier transformation of galaxy number counts in the Kaiser approximation, Eq. (\ref{Delta}), is given by 
\be
\! \langle \tilde \Delta(\bk, \bn, \bar{z}) \tilde \Delta (\bk',  \bn, \bar{z}) \rangle = (2 \pi)^3 P(\bk,  \bn, \bar{z}) \, \delta(\bk +\bk')\,,
\ee
where the power spectrum is given by (omitting the dependence on $ \bn, \bar{z}$) 
\be
\begin{split}\label{pp}
P(\bk) &= P^\text{iso}_\text{(s)} + P^\text{iso}_\text{(v)}  + P^\text{ani}_\text{(v)} \\ 
&= \big(b +  f (\bn\cdot \hat{\bk})^2\big)^2 P_{\delta \delta} (k) \\
&-\frac{k^2}{\HH^2} \omega (\bn \cdot \hat{\bk})^2  (1-(\bn\cdot \hat{\bk})^2)P_{\Omega}(k)\\
& -\frac{k^2}{\HH^2} (\bn\cdot \hat{\bk})^2 \hat n^i \hat n^j \omega_{ij}^A P_{\Omega}(k)\, ,
\end{split}
\ee
where all the isotropic contribution is in the first two lines and the anisotropic one, $ P^\text{ani}_\text{(v)}$,
is in the last line.%
 The tensor $\omega_{ij}^A$ is defined in Eq.\ (\ref{omegaA}). The isotropic contribution depends on directions only through the angle between the mode $\bk$ and the line-of-sight, i.e.\ it  can be expanded in a basis of Legendre polynomials as 
\be
P^{\text{iso}} (\bk, \bn, \bar{z}) = \sum\limits_\ell p_\ell(k, \bar{z}) \PP_\ell (\bn\cdot \hat{\bk})\,.
\label{isopk}
\ee
We observe that this is not a specific property of redshift-space distortions but simply a consequence of statistical isotropy. Hence Eq.~(\ref{isopk}) holds for all the relativistic contributions to the galaxy number counts.

When statistical isotropy is violated, we expand the power spectrum in terms of the orthonormal set of bipolar spherical harmonics, as 
\be
P^{\text{ani}}_\text{(v)} (\bk, \bn, \bar{z}) = \sum_{\ell \ell' \, J M} \, \pi_{\ell\ell'}^{JM}(\bar z, k ) X^{JM}_{\ell\ell'} (\hat \bk,\bn) \,,
\label{biposhexp}
\ee
where
\be
\begin{split}
X^{JM}_{\ell\ell'} (\hat \bk,\bn) = \sum_{mm'} \mathbf{C}_{\ell m \ell'm'}^{JM} Y_{\ell m} (\hat \bk) Y_{\ell'm'}(\bn) \,.
\end{split}
\ee
In the case where there is not anisotropic signal in the power spectrum, the coefficients $\pi_{\ell\ell'}^{JM}$ simply reduce to 
\be
\pi_{\ell\ell'}^{JM} = \frac{4 \pi}{\sqrt{2\ell+1}} p_\ell(k) \delta_{J,0} \delta_{M,0} \delta_{\ell, \ell'}\,,
\ee
and we recover the expansion of Eq.~(\ref{isopk}).

For convenience, we can split the anisotropic contribution to the power spectrum (\ref{pp}) in three contributions
\be
\begin{split}
P^{\text{ani}}_\text{(v)} (\bk) &= -\frac{k^2}{\HH^2} (\bn\cdot \hat\bk)^2 \hat n^i \hat n^j \omega_{ij}^A P_{\Omega}(k) \\&= P^{\text{(a)}} (\bk)  +P^{\text{(b)}} (\bk)  +P^{\text{(c)}} (\bk) 
\end{split}
\ee
where we have separated the three cases:
\begin{enumerate}
\item[(a)] $\omega_{ij}^A = \omega_{ij}$\,,
\item[(b)] $\omega_{ij}^A= -\omega_{il}\hat{k}^{l}\hat{k}_{j}-\omega_{lj}\hat{k}^{l}\hat{k}_{i}$\,,
\item[(c)] $\omega_{ij}^A =\omega_{lm}\hat{k}^{l}\hat{k}^{m}\hat{k}_{i}\hat{k}_{j}$\,,
\end{enumerate}
so that
\begin{align}
&P^{\text{ani}}_\text{(v)}  (\bk, \bn, \bar{z})\nn\\
&= \! \sum_{\ell \ell' \, J M} \! \big(\pi_{\ell\ell'}^{JM\text{(a)}} +\pi_{\ell\ell'}^{JM\text{(b)}} +\pi_{\ell\ell'}^{JM\text{(c)}} \big) X^{JM}_{\ell\ell'} (\hat \bk,\bn) \,.
\label{biposhexp-2}
\end{align}
Note that this splitting has no direct physical interpretation: each contribution has a scalar component which however disappears in the sum of Eq.~(\ref{biposhexp-2}).
These contributions can be written in terms of the eigenvectors and eigenvalues $\hat\om^I$ and $\la_I$. After a long but straightforward computation we find 
\begin{widetext}
\be
\pi_{\ell\ell'}^{JM\text{(a)}}  = -\frac{16 \pi^{3/2}}{45} \frac{k^2}{\HH^2} P_{\Omega}(k)\sum\limits_I \lambda_I Y^*_{2M}(\hat \omega_I)  \left( \delta_{\ell,0}\delta_{\ell',2} +2\sqrt{\frac{2\ell'+1}{5}} \tj{2}{2}{\ell'}{0}{0}{0}\delta_{\ell,2} \right)\delta_{J,2}  \,,
\label{coeffa}
\ee
\be
\begin{split}
\pi_{\ell\ell'}^{JM\text{(b)}}  =& -\frac{16 \pi^{3/2}}{5} \frac{k^2}{\HH^2} P_{\Omega}(k)\sqrt{(2\ell+1)(2\ell'+1)}\sqrt{\frac{2}{15}}\sum\limits_I \lambda_IY^*_{2M}(\hat \omega_I) \\
&\times \bigg[ 2 \tj{3}{1}{\ell}{0}{0}{0}  \tj{3}{1}{\ell'}{0}{0}{0} \ninej{1}{2}{1}{\ell}{3}{\ell'} + 3 \tj{1}{1}{\ell}{0}{0}{0}  \tj{1}{1}{\ell'}{0}{0}{0} \ninej{1}{2}{1}{\ell}{1}{\ell'} \bigg] \delta_{J,2}\,,
\end{split}
\label{coeffb}
\ee
\be
\begin{split}
\pi_{\ell\ell'}^{JM\text{(c)}}  =&  -\frac{16 \pi^{3/2}}{15} \frac{k^2}{\HH^2} P_{\Omega}(k)\sum\limits_I \lambda_I Y^*_{2M}(\hat \omega_I) \bigg[ \frac{1}{5} \delta_{\ell,2}\delta_{\ell',0}+ \frac{8}{105} \sqrt{2\ell+1}  \tj{4}{2}{\ell}{0}{0}{0} \delta_{\ell',4} \\
& + \frac{4}{7 \sqrt{5}}  \sqrt{2\ell+1}  \tj{2}{2}{\ell}{0}{0}{0} \delta_{\ell',2} \bigg] \delta_{J,2} \,,
\end{split}
\label{coeffc}
\ee
\end{widetext}
where curly brackets $\{ \}$ denote the Wigner 6j-symbols, see e.g.~\cite{Messiah}.
We first note that vector anisotropies can only generate $J=2$ modes. This is not surprising as they are the product of two $j=1$ states which can give either $J=0$ which is isotropic or $J=2$. The
 triangular relation imposed by the 3j and 6j symbols determines the limits of the sum in the expansion in Eq.~(\ref{biposhexp-2}). It is easy to see that both $\ell$ and $\ell'$ have to be even and, more precisely, in $\{0,2,4,6\}$. 
We can now  reconstruct the correlation function (\ref{corr-def}) from the power spectrum. This is similar to the isotropic case in which the Fourier- and real-space coefficients in the Legendre expansion are related by
\be
\xi_\ell (x) = i^\ell \int \frac{k^2 d k }{2 \pi^2} j_\ell(k x)  p_{\ell}( k ) \,.
\ee
Explicitly, the coefficients of the BipoSH expansion of the correlation function, Eq. (\ref{corrB}), are related to the ones of the power spectrum, Eq. (\ref{biposhexp})  by 
\be
\xi_{\ell\ell'}^{JM} (x) = i^\ell \int \frac{k^2 d k }{2 \pi^2} j_\ell(k x)  \pi_{\ell\ell'}^{JM}( k ) \,.
\ee
With this we can rewrite the real-space version of Eqs.~(\ref{coeffa})-(\ref{coeffc}) in terms of the  $C_{n}^{\Omega}$, which we defer to an appendix: Eqs.~(\ref{realcoeffa})-(\ref{realcoeffc}). The sum of the three contributions can be cast in matrix form as (remember all terms with $J\neq 2$ vanish) 
\begin{widetext}
\be
\big(\xi^{2M}_{\ell\ell'}  \big)= \frac{16 \pi^{3/2}}{5}\sum\limits_I \lambda_I Y^*_{2M}(\hat \omega_I) 
\begin{pmatrix}
 0 & 0 & \frac{1}{35} C^\Omega_0 & 0 & 0 & 0 &0 \\
 0 & 0 & 0 & 0 & 0 & 0 &0 \\
  \frac{1}{25} C^\Omega_2 & 0 & -\frac{1}{5} \sqrt{\frac{2}{35}}C^\Omega_2 & 0 & \frac{2}{225} \sqrt{\frac{2}{7}}C^\Omega_2 & 0 &0 \\
   0 & 0 & 0 & 0 & 0 & 0 &0 \\
    0 & 0 & 0 & 0 &  -\frac{4}{9\sqrt{385}} C^\Omega_4 & 0 &0 \\
   0 & 0 & 0 & 0 & 0 & 0 &0 \\  
    0 & 0 & 0 & 0 &  -\frac{8}{63\sqrt{55}} C^\Omega_6 & 0 &0
\end{pmatrix} \,,
\label{bigM}
\ee
\end{widetext}
where $C^\Omega_\ell = C_{\ell}^{\Omega}(z,x)$. Equation~(\ref{bigM}) is one of the main results of this paper. It shows in complete generality that {\it any} anisotropic signal induced by redshift-space distortion in the galaxy correlation function is encoded in the functions $\xi_{\ell\ell'}^{2M}$ (which depend in principle on redshift and on galaxy separation). The six non-zero coefficients $\boldsymbol{\xi} = \{\xi^{2M}_{02},\xi^{2M}_{20},\xi^{2M}_{22},\xi^{2M}_{24},\xi^{2M}_{44},\xi^{2M}_{64} \}$ are therefore the equivalent of the monopole, quadrupole and hexadecapole that are measured in standard redshift surveys, when anisotropies are assumed to be absent. As we will show below, these six coefficients can be directly extracted from catalogs of galaxies, by averaging over pairs of galaxies with an appropriate weighting. A detection of a non-zero $\xi_{\ell\ell'}^{2M}$ would represent a smoking gun for the presence of anisotropies in the galaxies peculiar velocities. Note that the dependence of the $\xi_{\ell\ell'}^{2M}$ on the model responsible for the anisotropies is encoded in the 
$C_\ell^\Omega(z,x)$ and in the eigenvectors $\hat\om_I$ and eigenvalues $\la_I$. In the following, we construct estimators for the six non-zero coefficients $\boldsymbol{\xi}$ and we forecast the detectability of these coefficients with future surveys. 

\begin{figure*}[th]
\includegraphics[width=15cm]{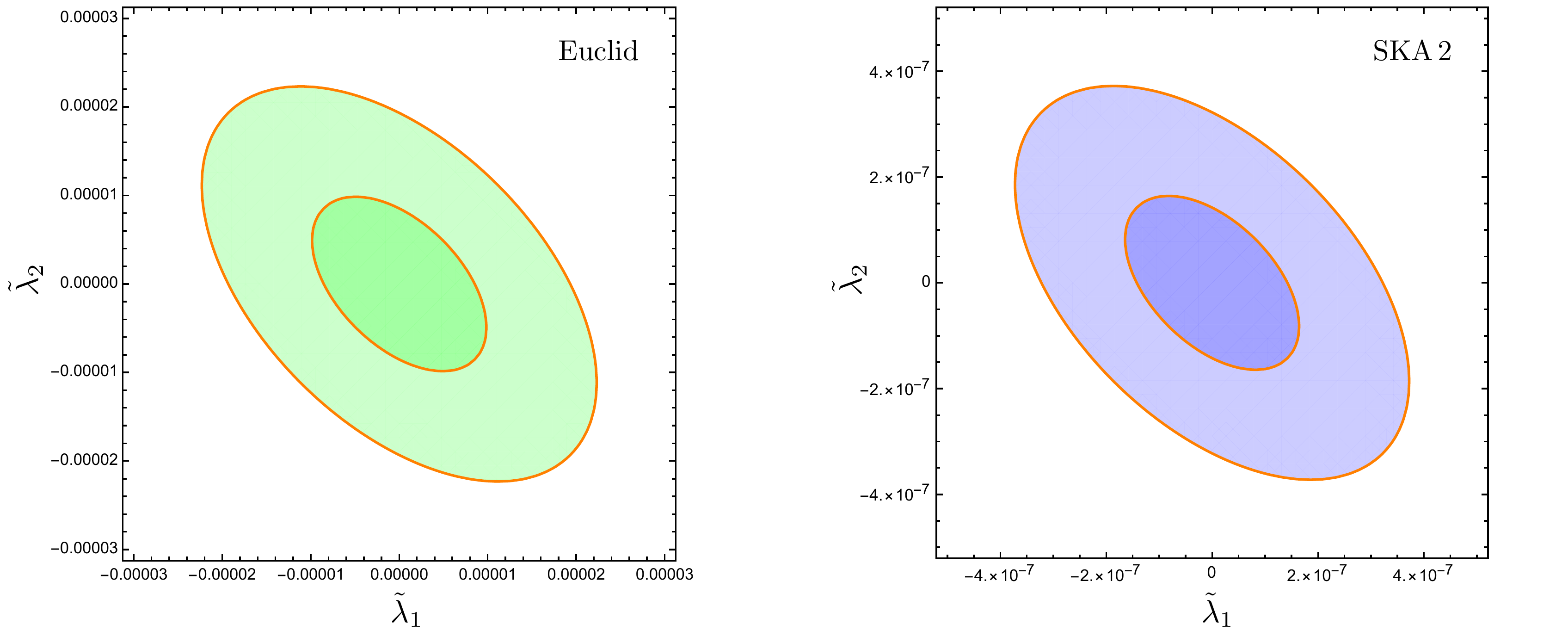}
\caption{\label{fisher} Constraints on amplitudes of anisotropies for the model (A). We have rescaled the parameters $\lambda_I$ as $\tilde{\lambda}_I\equiv A_V\lambda_I$. Compare this with the amplitude of the isotropic vorticity power spectrum generated by shell crossing in the standard cold dark matter scenario, where $A_{V,\text{iso}}\sim 10^{-5}$, see e.g.\ Ref.~\cite{Pueblas:2008uv}.}
\end{figure*}

\section{Forecast for LSS surveys\label{sec:fisher}}
We now forecast the constraints on the anisotropy parameters -- which we define later -- as expected from future redshift surveys. In the next section we define our estimators for the BipoSH coefficients and compute their covariance matrix.
  
\subsection{Estimator \& covariance}
To estimate the expansion coefficients the obvious choice is to weight the correlation function by $X_{\ell\ell'}^{2M}$, in the same way that we weight the two-point function by the Legendre polynomials $\mathcal{P}_\ell$ to estimate the multipoles. In a binned survey the estimator is
\be
\hat \xi_{\ell\ell'}^{\,2M} (x) = a_N \sum\limits_{i,j} \Delta_i \Delta_j X_{\ell\ell'}^{2M*} (\hat x_{ij}, \hat n_{ij})\delta_K(x_{ij}-x) \,,
\ee
where $\delta_K$ is the Kronecker delta, $\Delta_i$ the galaxy over-density in the bin labelled by the index $i$ and we have defined $\bx_{ij} = \bx_i-\bx_j$, $\mathbf{n}_{ij}= 1/2(\bx_i+\bx_j)$. The normalisation factor $a_N$ is found by imposing that the estimator is unbiased, 
\be
\left\langle \hat \xi_{\ell\ell'}^{\,2M} \right\rangle =  \xi_{\ell\ell'}^{\,2M} \,,
\ee
in the continuous limit
\be
\sum\limits_i \rightarrow \frac{1}{L_p^3} \int d^3 x_i \,,\quad  \delta_K(x_{ij}-x) \rightarrow L_p \delta_D(x_{ij}-x)\,,
\ee
where $L_p$ denotes the pixel size and $V$ is the total volume of the survey. We obtain 
\be
a_N = \frac{3 L_p^5}{Vx^2}\,.
\ee
We also have $a_N=1/N(x)$, where $N(x)$ is the number of pixels which contribute to the estimator. 
The variance of the estimator is defined as 
\be
\text{var}\left(  \hat \xi_{\ell\ell'}^{\,2M} \right) \equiv \text{var}^M_{\ell\ell'} =\left\langle \left(  \hat \xi_{\ell\ell'}^{\,2M} \right)^2 \right\rangle -\left\langle \hat \xi_{\ell\ell'}^{\,2M} \right\rangle^2\,,
\ee
and we recall that $\langle \Delta_i \Delta_j \rangle$ contains a Poisson noise contribution and a cosmic variance (CV) contribution,
\be
\langle \Delta_i \Delta_j \rangle = \frac{1}{d \bar n}\delta_{ij} + C_{ij}^\Delta \,,
\ee
where $d \bar n$ is the mean number of galaxies per pixel. The correlation $C_{ij}^\Delta$ is due both to the scalar and vector parts of $\Delta$. However, the scalar component strongly dominates over the vector one, so that we can neglect the latter. Physically, this reflects the fact that even though the coefficients $\xi_{\ell\ell'}^{2M}$ are constructed to remove the scalar isotropic signal and to isolate the vector anisotropic signal, the covariance of these coefficients is still affected (and dominated) by the scalar contribution. We then obtain three different contributions to the variance which are understood respectively as the Poisson term (P), the mixed term (M) and the CV term (C). Explicitly, we find
\begin{align} 
\text{var}_P (x,x')&=\frac{6 V}{x^2 N^2_\text{tot}} \delta_D(x-x')\,,\\
\text{var}_M(x,x') &= \frac{24}{\pi N_\text{tot}} \int dk \, k^2 P(k,\bar z) j_\ell(k x) j_\ell(kx')\nn\\
&\quad  \times\sum\limits_w c_w \beta^w_{\ell\ell'}\,,\\
\text{var}_C (x,x') &= \frac{12}{\pi V} \int dk \, k^2 P^2(k,\bar z) j_\ell(k x) j_\ell(kx') \nn\\
&\quad\times \sum\limits_\sigma \tilde c_\sigma \beta^\sigma_{\ell\ell'}\,,
\end{align}
where $N_\text{tot}$ is the total number of tracers in the catalog and the indices $w,\sigma$ take values $w=0, 2, 4$ and $\sigma=0, 2, 4, 6, 8$. The explicit form of the coefficients $\beta^\sigma_{\ell\ell'}$ and details on the derivation of the various contributions of the variance can be found in appendix~\ref{sec:covariance}, where we also compute the covariance matrix of the estimator, defined as 
\be\label{cog}
\begin{split}
\text{cov}&\left(  \hat \xi_{\ell_1\ell_1'}^{\,2M_1} ,\hat \xi_{\ell_2\ell_2'}^{\,2M_2} \right) \equiv \text{cov}^{M_1M_2}_{\ell_1\ell_1'\ell_2\ell_2'}   \\& =\left\langle  \hat \xi_{\ell_1\ell_1'}^{\,2M_1}\hat \xi_{\ell_2\ell_2'}^{\,2M_2} \right\rangle -\left\langle \hat \xi_{\ell_1\ell_1'}^{\,2M_1} \right\rangle \left\langle \hat \xi_{\ell_2\ell_2'}^{\,2M_2}  \right\rangle \,.
\end{split}
\ee

\subsection{Fisher forcasts}\label{sub:fisherforcast}
We now want to forecast the constraints on the anisotropic parameters from a survey. Given a model for the anisotropy power spectrum, i.e. a parametrization for $P_\Omega$, we are left with the 5 degrees of freedom (d.o.f.) of the symmetric traceless tensor $\omega_{ij}$ and an overall amplitude $A_V$ for the vector power spectrum, which can be reabsorbed in a  redefinition of $\omega_{ij}$. Following our decomposition in Eq.~(\ref{wdec}) we identify the d.o.f.\ as the eigenvalues and eigenvectors of $\omega_{ij}$. On the one hand the eigenvalues are of zero-sum so that we can pick the first two $\lambda_1, \lambda_2$ as independent and the third one is fixed to $-(\lambda_1 + \lambda_2)$. On the other hand we find it convenient to parametrize the three orthonormal eigenvectors $\hat{\omega}_I$ in terms of the three angles of an Euler-rotation which rotates the canonical basis of $\mathbb{R}^3$ into the  $\hat{\omega}_I$, 
\be
\hat{\omega}_I \equiv R(\alpha, \beta, \gamma) \cdot \hat{e}_I\,,
\label{eulwang}
\ee
where $\hat{e}_I$ are the three orthonormal vectors  of $\mathbb{R}^3$ and $R(\alpha, \beta, \gamma)$ is the rotation matrix with Euler angles $\alpha, \beta, \gamma$. Furthermore we can absorb the amplitude $A_V$ in the eigenvalues by defining $\tilde \lambda_I = A_V \lambda_I$. In summary the 5 d.o.f.\ of the tensor $\bar\omega_{ij}$ and the overall amplitude $A_V$ are encoded in our parameter space
\be
\theta =\{\tilde \lambda_1, \tilde \lambda_2, \alpha, \beta,\gamma \} \,.
\ee

The  Fisher matrix is defined as
\begin{align}\label{Fisher}
F_{\theta\theta'} \equiv \frac{1}{2} \frac{\pd^2 \chi^2}{\pd \theta \pd \theta'} = \sum\limits_{\mathcal{A},\mathcal{A}'} \frac{\pd \langle \hat \xi_\mathcal{A} \rangle}{\pd \theta} \bigg|_\text{f} \text{cov}^{-1}_{\mathcal{A}\mathcal{A}'}  \frac{\pd \langle \hat \xi_\mathcal{A'}^*\rangle}{\pd \theta'} \bigg|_\text{f}
\end{align}
where, schematically, $\mathcal{A} = \{\ell_1,\ell_1',M_1,x_i,z_1 \}$, $\mathcal{A'} = \{\ell_2,\ell_2',M_2,x_j,z_2\}$, and the derivatives are evaluated at the fiducial model. The Fisher matrix contains therefore a sum over the six non-zero coefficients which constitute our data, over all pixels separations $x_i, x_j$ and over all bins of redshifts $z_i, z_j$. The covariance matrix properly accounts for all correlations between these quantities, except for the correlations between different redshift bins $z_i\neq z_j$, which we assume to be uncorrelated, since the bin size that we consider is sufficiently large.  We then have
\be
 \text{cov}_{\mathcal{A}\mathcal{A}'} =  \text{cov}_{\ell_1\ell_1'\ell_2\ell_2'}(x_i,x_j) \de_{M_1M_2} \de_{z_1z_2} \,.
 \label{covdelta}
\ee
We recall that, according to the Cramer-Rao inequality, the Fisher matrix provides a lower bound on the marginal parameter uncertainty $\sigma_\theta$ as
\be
\sigma_\theta \geqslant \sqrt{(F^{-1})_{\theta\theta}} \,.
\ee
\begin{figure*}[th]
\includegraphics[width=10.5cm]{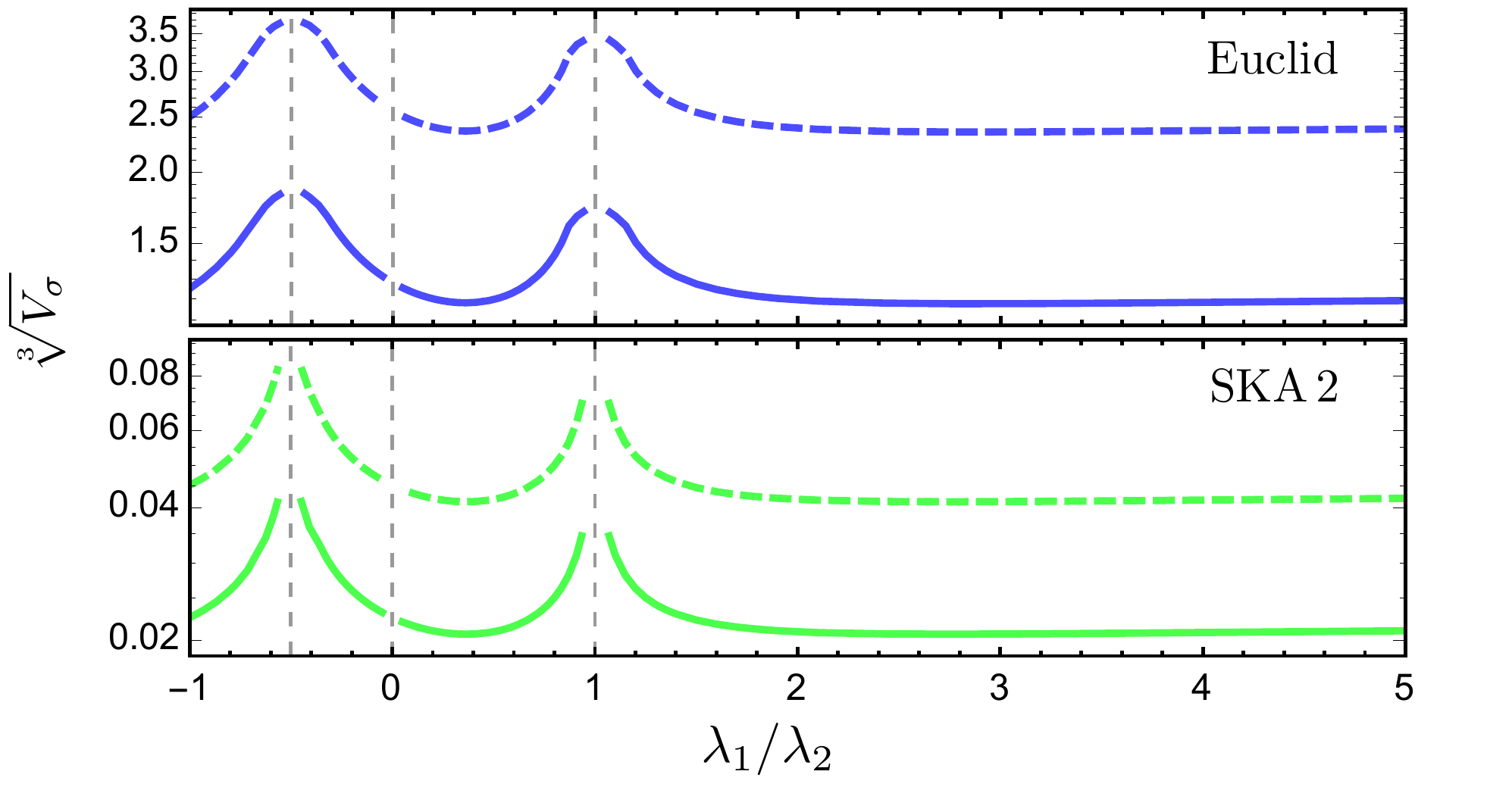}
\caption{\label{fishervol} Volume of the $1\sigma$ (solid) and $3\sigma$ (dashed) ellipsoids in the $\alpha-\beta-\gamma$ space as a function of the ratio between the $\lambda_A$ and $\lambda_B$. The anisotropic amplitude is set to $A_V = 10^{-5}$. The constraint should be compared to the the cube root of the Haar volume $\sqrt[3]{79}\sim 4$.} 
\end{figure*}
We start by constraining the parameters $\lambda_1, \lambda_2$. 
The sub-matrix is then written
\begin{widetext}
\be
\begin{split}\label{fish}
F_{\tilde \la_A\tilde \la_B} &= \sum\limits_{\{ z_\text{bin} \}} \sum\limits_{i,j} \sum\limits_{\ell_1\ell_1'\ell_2\ell_2'} \sum\limits_M \frac{\pd \xi^M_{\ell_1\ell_1'}(x_i,z)}{\pd \tilde \la_A} \text{cov}^{-1}_{\ell_1\ell_1'\ell_2\ell_2'}(x_i,x_j)  \frac{\pd \xi^{M*}_{\ell_2\ell_2'}(x_j,z)}{\pd \tilde \la_B} \\
& = \sum\limits_{\{ z_\text{bin} \}} \sum\limits_{i,j} \sum\limits_{\ell_1\ell_1'\ell_2\ell_2'} \frac{5}{4\pi}\left(2+\PP_2(\de_{AB})\right) \tilde \xi_{\ell_1\ell_1'}(x_i,z) \text{cov}^{-1}_{\ell_1\ell_1'\ell_2\ell_2'}(x_i,x_j)  \tilde \xi^*_{\ell_1\ell_1'}(x_j,z)\,,
\end{split}
\ee
\end{widetext}
where we have defined
\be
\xi^{2 M}_{\ell_1\ell_1'}\equiv  A_V \sum_I \lambda_I Y_{2 M}^*(\hat{\omega}_I)\tilde{\xi}_{\ell_1\ell_1'}\,,
\ee
by explicitly writing the amplitude $A_V$ of $P_\Omega$ out of the $C_\ell^\Omega$. We normalize this amplitude such that $\la_\text{max}\equiv 1$. The variables which determine the anisotropy are then the amplitude $A_V$, the ratio $\la_2/\la_1=\la_2$ and the three angles $(\alpha,\beta,\gamma)$ which determine the orientation. 
For the second equal sign of Eq.~(\ref{fish}) we have performed the sum over $M$ 
using that 
\be
\frac{\partial\xi^{2 M}_{\ell_1\ell_1'}}{\partial\tilde \lambda_A} =Y_{2 M}^*(\hat{\omega}_A)\tilde{\xi}_{\ell_1\ell_1'}-Y_{2 M}^*(\hat{\omega}_3)\tilde{\xi}_{\ell_1\ell_1'}\,,
\ee
together with the orthogonality properties of products of spherical harmonics. 
We observe that the final result does not depend on the fiducial values of the parameters $\tilde \lambda_I$ since they enter linearly in the estimator $\hat\xi_{\ell_1\ell_2}^{2M}$. We also note that we did not need to fix any fiducial direction $\hat{\omega}_I$ since the dependence on $\hat{\omega}_I$ cancels out in the final result.

In appendix \ref{den} we show that the off-diagonal blocks of the full Fisher matrix (\ref{Fisher}) are vanishing, hence $F_{\theta\theta'}$ has a block diagonal structure 
\be\label{block}
[F_{\theta\theta'}] =\left[
\begin{array}{c|c}
F_{\lambda_A\lambda_B}& \mathbf{0}\\
\hline
\mathbf{0}& F_{\alpha\beta\gamma}
\end{array}
\right] \,.
\ee

As a consequence of the block structure of the Fisher matrix, it follows that the constraints on the amplitudes $\lambda_I$ can be derived directly with Eq.~(\ref{fish}). In particular, they do not depend on the fiducial values of the eigenvectors $\hat \omega_I$. This reflects the fact that the precision with which we can measure the eigenvalues does not depend on the direction of the anisotropy. The constraints on directions, i.e.  $(\sigma_\alpha, \sigma_\beta, \sigma_\gamma)$, can be obtained by inverting the lower block of the Fisher matrix. It turns out that the constraints on each of the directions depend on the fiducial values of both the eigenvalues and the eigenvectors of $\bar \omega_{ij}$. However, this direction dependence is somewhat artificial, as we could have chosen our basis directions differently. Instead of considering each direction independently, it makes more sense to compute the volume of the ellipsoid described by the constrains on ($\alpha$, $\beta$, $\gamma$), using the Haar measure $\di\mu= \sin\beta\di\alpha \di\beta \di\gamma$. Note that with this non-normalized Haar measure, the volume of the rotation group $SO(3)$ is $2(2\pi)^2\simeq 79$. We can think of this uncertainty volume  as the inverse of a `figure of merit' for the average accuracy with which we can recover the directional information. This combined direction constraint has the great advantage that it does not depend on the fiducial model for the directions, but only on the choice of the eigenvalues' ratio $\lambda_A/\lambda_B$ and the vector amplitude $A_V$. This remaining dependence is physical and simply reflects the fact that the precision with which we can measure the direction of the anisotropy does obviously depend on how large it is.

\subsection{A model for vector perturbations}

To illustrate how our general formalism can be used, we consider an explicit model in which a non-isotropic vector contribution to the galaxy peculiar velocities gives new contributions to the correlation function. We derive constraints on the directions and amplitudes of anisotropies for both a Euclid-like and SKA2-like survey. The specifications for these surveys are taken from \cite{2011arXiv1110.3193L} and \cite{Villa:2017yfg} respectively: the two redshift ranges are $z \in [ 0.7,2.0]$ for Euclid and $z \in [0.1,2.0]$ for SKA2 and we split them into 14 and 19 bins of thickness $\Delta z =0.1$ respectively, with $L_p = 2\, \text{Mpc}/h$. This choice of $L_p$ is motivated by the fact that this pixel size gives the best constraints in \cite{Bonvin:2017req}. 
Note that in the isotropic case, exploiting separations as small as $2\, \text{Mpc}/h$ does require a good understanding of the scalar non-linear signal at those scales, which is highly non-trivial. In the anisotropic case however, since the scalar part does not contribute to the estimators $\hat \xi_{\ell\ell'}^{\,2M}$, but only to the covariance, we can exploit very small separations even without a very precise modelling of the scalar behaviour at those scales. As maximum separation we choose 40 Mpc/h. \\

Until this point our formalism has been model independent but, clearly, to forecast the detectability of the anisotropy parameters we have to assume a shape for $P_\Omega(k)$. As an example we choose the isotropic vorticity power spectrum from N-body simulations while we note that, as we have stated before, the isotropic and anisotropic $P_\Omega$ can in principle be different.
 
According to the numerical simulations of Ref.~\cite{Pueblas:2008uv,Jelic-Cizmek:2018gdp}, the vorticity power spectrum appears to evolve as  $\Hcal(z)^2 f(z)^2 D_1(z)^7$ at large scales. At small scales, the evolution has an additional scale-dependence, leading to a suppression of power at small scales at late times, see Fig.\ 4 of~\cite{Pueblas:2008uv}. In the following we will ignore this small-scale dependence and assume that the power spectrum at redshift $z$ is given by\footnote{Note that the constraints obtained in this way are conservative, because we underestimate the vorticity power spectrum at small scales for large redshift.}
\begin{equation*}
P_\Omega (k,z)=P_\Omega (k,z=0)\left(\frac{\Hcal(z)f(z)}{\Hcal_0f(z=0)}\right)^2\left(\frac{D_1(z)}{D_1(z=0)}\right)^7\, .
\end{equation*}
We use the vorticity power spectrum plotted in Fig.\ 4 of~\cite{Pueblas:2008uv} to construct the following fit for $P_\Omega$,
\[
P_\Omega (k,z=0)=A_V \frac{(k/k_*)^{n_\ell}}{\left[ 1+ (k/k_*) \right]^{n_\ell+n_s}}\quad \big({\rm Mpc}/h)^3\, , \label{Pom1343}
\]
where the power at large scales is given by $n_\ell=1.3$, the power at small scales by $n_s=4.3$ and the transition scale by $k_*=0.7\,h/$Mpc. From Fig.\ 4 of~\cite{Pueblas:2008uv} we find that the predicted amplitude for $P_\Omega$ is $A_V=10^{-5}$. \\

In Fig.~\ref{fisher} we use this spectrum to estimate the constraints on the eigenvalues $\tilde \la_{1,2}$. Note that there is no dependence on the fiducial values of the parameters $\tilde \lambda_I$ since they enter linearly in the estimator $\hat\xi_{\ell_1\ell_2}^{2M}$. Furthermore the constraints do not depend on the orientation of the eigenvectors due to the block diagonal structure of the Fisher matrix. The $1\sigma$-constraints on the amplitude of the eigenvalues are  $\sigma_\lambda\simeq  6\times 10^{-6}$ with Euclid and even $\sigma_\lambda \simeq 1\times 10^{-7}$ with SKA2. It is also interesting to note that the constraints are better if both eigenvalues have the same sign. This is of course owing to the fact that then the norm of the third eigenvalue is larger. With the SKA and optimistic assumptions we should therefore be able to constrain an anisotropic vector signal with amplitude of 1\% of the amplitude of the vorticity generated by shell-crossing in cold dark matter $A_{V,\text{iso}V}\sim 10^{-5}$ \cite{Pueblas:2008uv}. \\

In Fig.~\ref{fishervol} we show the volume of the ellipsoid described by the constraints on $(\alpha, \beta, \gamma)$. As we discussed above the constraint does not depend on the fiducial directions but  it depends on the fiducial values of $\tilde \lambda_{1,2}$ or, equivalently, on the choice of $A_V$ and the ratio $\lambda_1/\lambda_2$.  In the plot we fix the \textit{biggest} eigenvalue to $\lambda_\text{max} = A_V = 10^{-5}$. The features in the plot can be explained intuitively as follows. We first note that the constraint asymptotes to a constant for $\lambda_1/\lambda_2 >1$: this is a result of two concurrent effects. On one hand, as we keep the largest eigenvalue,  $\lambda_1$, fixed, the other, $\lambda_2$, becomes smaller, reducing the overall signature of the anisotropy. On the other hand, as the ratio increases, the departure from isotropy is more pronounced yielding better constraints.  Note that we could have fixed the \textit{smallest} eigenvalue equal to $A_V$: in this case as the ratio becomes bigger the overall signature of anisotropy increases and the two effects add up to give better constraints. Secondly the constraints are worst for $\lambda_1/\lambda_2=1$ or $-1/2$. In both cases this is because we approach a degeneracy: $\lambda_1 = \lambda_2$ or $\lambda_1 = \lambda_3$ respectively. Note that the constraints are slightly better in $\lambda_1 = \lambda_2$ w.r.t. the second case as the overall amplitude is bigger in this case. The absolute values  of the volume show that Euclid constrains the direction of the anisotropy only loosely, while the constraints from SKA2 are excellent, for our choice of amplitude $A_V = 10^{-5}$. Note that the constraint on the volume scales as $A_V^3$, so that decreasing $A_V$ by an order of magnitude would degrade the bounds in Fig.~\ref{fishervol} by a factor 10.

\section{Conclusions}

In this paper we have discussed the effects of an anisotropic vector component in the peculiar velocity field, focusing on the redshift-space distortions induced in the galaxy correlation function. We have presented a general method to isolate the anisotropic signal through a decomposition in bipolar spherical harmonics. We provide an analytical expression for the coefficients of this expansion which does not require the adoption of a specific model. We then show how one can practically use this approach to forecast constraints on the anisotropic sector for two upcoming redshift surveys. 

We derive two types of constraints, both on the total {amplitude} of the anisotropy and on the preferred direction (in terms of the $SO(3)$ volume of its Euler angles). We can compare our results with the constraints found in \cite{Bonvin:2017req} for the isotropic case which of course has no preferred direction. Given the block-diagonal form of the Fisher matrix, we find that we are able to achieve similar constraints on the amplitude of the vector modes (since we also assume the same shape for the spectrum). Let us however note that even though the constraints are similar, the interpretation of the result in the anisotropic case is cleaner, since in this case the scalar degrees of freedom do not contribute to the estimators and therefore they do not need to be accurately modelled.

This work is meant as a study of the feasibility of detecting an anisotropic vector signal in the galaxy two-point function and together with the analysis carried out in \cite{Bonvin:2017req}, it represents a comprehensive study of the detectability of vector modes in the correlation function. 
 
Given a model for the anisotropy,  one needs to determine not only the eigenvalues and the directions of its eigenvectors, but also the corresponding vector power spectrum. Here we just assumed this to be given by the vorticity spectrum generated by non-linear structure formation. This corresponds to a model where an anisotropy only affects the direction but not the strength of the generated vorticity, which is of course not true in general. In full generality the power spectrum could be reconstructed from the data as a function of multipole and redshift, at the price of much larger error bars. 

\newpage

\begin{acknowledgments}
C.B., R.D., M.K. and V.T.~acknowledge
funding by the Swiss National Science Foundation. G.C. acknowledges financial support from
ERC Grant No:  693024 and Beecroft Trust. I.S. is supported by the European
Regional Development Fund and the Czech Ministry of Education, Youth
and Sports (MŠMT) (Project CoGraDS — CZ.02.1.01/0.0/0.0/15\_003/0000437). \end{acknowledgments}

\newpage

\appendix

\begin{widetext}
\section{Covariance matrix}
\label{sec:covariance}

The variance of the estimator  $\xi_{\ell\ell'}^{2M}$  is given by
\be
\begin{split}
\text{var}\left(  \hat \xi_{\ell\ell'}^{\,2M} \right) 
& =a_N^2 \sum\limits_{ij} \sum\limits_{km} \langle \Delta_i \Delta_j \Delta_k \Delta_m \rangle X_{\ell\ell'}^{2M} (\hat x_{ij}, \hat n_{ij}) X_{\ell\ell'}^{2M*} (\hat x_{km}, \hat n_{km}) \delta_K(x_{ij}-x) \delta_K(x_{km}-x') \\
&= \text{var}_P+ \text{var}_M+ \text{var}_C \,.
\end{split}
\ee
Since $\langle \Delta_i \Delta_j \rangle$ contains a Poisson noise contributions and a cosmic variance (CV) contribution
\be
\langle \Delta_i \Delta_j \rangle = \frac{1}{d \bar n} \delta_{ij}+ C_{ij}^\Delta \,,
\ee
where $d \bar n$ is the mean number of galaxies per pixel, the three different contributions to the variance are understood respectively as the Poisson term, the mixed term and the CV term. The first terms is easily found
\be
\begin{split}
\text{var}_P(x,x') &= \frac{18 L_p^{10}}{V^2 (xx')^2} \frac{1}{d\bar n^2} \sum\limits_{ij} \sum\limits_{km} \delta_{ik} \delta_{jm}X_{\ell\ell'}^{2M} (\hat x_{ij}, \hat n_{ij}) X_{\ell\ell'}^{2M*} (\hat x_{km}, \hat n_{km}) \delta_K(x_{ij}-x) \delta_K(x_{km}-x') \\
&= \frac{6 V}{x^2 N^2_\text{tot}} \delta_D(x-x') \,,
\end{split}
\ee
where we have set the factor $(1+(-1)^\ell)=2$ as only even $\ell$ appear in the expansion. The mixed term is 
\begin{align}
\text{var}_M(x,x') &= \frac{18 L_p^{10}}{V^2 (xx')^2} \frac{1}{d\bar n} \sum\limits_{ij} \sum\limits_{km} \big(\delta_{ik} C_{jm}^\Delta +\delta_{jm} C_{ik}^\Delta \big)X_{\ell\ell'}^{2M} (\hat x_{ij}, \hat n_{ij}) X_{\ell\ell'}^{2M*} (\hat x_{km}, \hat n_{km}) \nonumber  \delta_K(x_{ij}-x) \delta_K(x_{km}-x')  \\
&= \frac{18 L_p^{10}}{V^2 (xx')^2} \frac{2}{d\bar n} \sum\limits_{ij} \sum\limits_{m} C^\Delta_{jm} X_{\ell\ell'}^{2M} (\hat x_{ij}, \hat n_{ij}) X_{\ell\ell'}^{2M*} (\hat x_{im}, \hat n_{im})  \delta_K(x_{ij}-x) \delta_K(x_{im}-x') \nonumber \,.
\end{align}
We use the flat-sky expression for $C_{ij}^\Delta$
\be
C_{ij}^\Delta(\bar z) = \frac{1}{(2\pi)^3} \int d^3k \,\, e^{i \bk \cdot(\bx_j-\bx_i)} P(k,\bar z) \Big(c_0 \PP_0\ndotk +c_2 \PP_2\ndotk +c_4 \PP_4\ndotk \Big) \,,
\label{FSxi}
\ee
and we perform (in the continuous limit) the following change of variables $\by_j = \bx_j -\bx_i$, $\by_m = \bx_m -\bx_i$ together with $\bx_i = \bn$. We obtain
\be 
\text{var}_M(x,x') = \frac{24}{\pi N_\text{tot}} \int dk \, k^2 P(k,\bar z) j_\ell(k x) j_\ell(kx') \Big(c_0 \beta^0_{\ell\ell'}+c_2 \beta^2_{\ell\ell'}+c_4\beta^4_{\ell\ell'} \Big)\,,
\label{varM}
\ee
where we have defined the coefficients 
\be
\beta^\sigma_{\ell\ell'} = (2\ell+1)(2\ell'+1) \tj{\sigma}{\ell}{\ell}{0}{0}{0} \tj{\sigma}{\ell'}{\ell'}{0}{0}{0} \ninej{\ell'}{\ell}{2}{\ell}{\ell'}{\sigma} \,.
\ee
Finally the CV term is given by
\be
\begin{split}
\text{var}_C (x,x') &= \frac{18 L_p^{10}}{V^2 (xx')^2} \sum\limits_{ij} \sum\limits_{km} C_{jm}^\Delta C_{ik}^\Delta X_{\ell\ell'}^{2M} (\hat x_{ij}, \hat n_{ij}) X_{\ell\ell'}^{2M*} (\hat x_{km}, \hat n_{km})  \delta_K(x_{ij}-x) \delta_K(x_{km}-x')\,,
\end{split}
\ee
we can perform a similar change of variable as above $\by_j = \bx_j -\bx_i$, $\by_m = \bx_m -\bx_k$ so that, after substituting Eq.~(\ref{FSxi}) twice, the two exponentials are written
\be
e^{i \bk \cdot(\bx_m-\bx_j)}  e^{i \bk' \cdot(\bx_k-\bx_i)} \rightarrow e^{i \bk \cdot(\by_m-\by_j)}  e^{i (\bk+\bk') \cdot(\bx_k-\bx_i)}\,,
\ee
and the integral over $\bx_k$ enforces $\bk=-\bk'$.  The angular integrals are performed with the properties of BiPoSH as before and we obtain
\be 
\text{var}_C (x,x') = \frac{12}{\pi V} \int dk \, k^2 P^2(k,\bar z) j_\ell(k x) j_\ell(kx') \sum\limits_\sigma \tilde c_\sigma \beta^\sigma_{\ell\ell'}\,,
\label{varCV}
\ee
where
\begin{align}
\tilde c_0 &= c_0^2 +\frac{c_2^2}{5}+\frac{c_4^2}{9} \,,\\
\tilde c_2 &= \frac{2}{7}c_2 (7c_0+c_2) +\frac{4}{7} c_2 c_4+\frac{100}{693}c_4^2 \,, \\
\tilde c_4 &= \frac{18}{35} c_2^2 +2 c_0 c_4 +\frac{40}{77} c_2 c_4 +\frac{162}{1001} c_4^2 \,,\\
\tilde c_6 &= \frac{10}{99} c_4 (9 c_2 +2 c_4) \,,\\
\tilde c_8 &= \frac{490}{1287} c_4^2 \,.
\end{align}
The computation for the off-diagonal covariance matrix, defined in Eq. (\ref{cog}), 
follows the same steps with the exception that Poisson noise does not contribute for off-diagonal components as it is proportional to $\delta_{\ell_1\ell_2}\delta_{\ell'_1\ell'_2} \delta_{M_1 M_2}$. Furthermore the mixed and Cosmic contributions are proportional to $\delta_{M_1 M_2}$ and the general case is obtained from Eqs.~(\ref{varM}) and (\ref{varCV}) by substituting the product of spherical Bessel functions inside the integral with $j_{\ell_1}(x) j_{\ell_2} (x')$ and redefining the $\beta$ coefficients as
\be
\beta^\sigma_{\ell_1\ell_1'\ell_2\ell'_2} = i^{\ell_2-\ell_1} \sqrt{(2\ell_1+1)(2\ell_1'+1)(2\ell_2+1)(2\ell_2'+1)} \tj{\sigma}{\ell_1}{\ell_2}{0}{0}{0} \tj{\sigma}{\ell_1'}{\ell_2'}{0}{0}{0} \ninej{\ell_1'}{\ell_1}{2}{\ell_2}{\ell_2'}{\sigma} \,.
\ee

\section{Fisher matrix}\label{den}

In this appendix we sketch a proof of why the off-diagonal blocks of the Fisher matrix (\ref{block})  vanish, i.e. $F_{\lambda_A, \alpha_i}=0$. We have 
\begin{align}\label{Fisher1}
F_{\lambda_A, \alpha_i}&=\sum\limits_{\{ z_\text{bin} \}} \sum\limits_{i,j} \sum\limits_{\ell_1\ell_1'\ell_2\ell_2'} \sum\limits_M \frac{\pd \xi^M_{\ell_1\ell_1'}(x_i,z)}{\pd \lambda_A} \text{cov}^{-1}_{\ell_1\ell_1'\ell_2\ell_2'}(x_i,x_j)  \frac{\pd \xi^{M*}_{\ell_2\ell_2'}(x_j,z)}{\pd \alpha_i} \\
&=\sum_M\left(Y^*_{2 M}(\hat{\omega}_A)-Y^*_{2 M}(\hat{\omega}_3)\right)\frac{\partial}{\partial \alpha_i}\left(\sum_I \lambda_I Y_{2 M}(\hat{\omega}_I)\right) \sum\limits_{\{ z_\text{bin} \}} \sum\limits_{i,j} \sum\limits_{\ell_1\ell_1'\ell_2\ell_2'} \tilde \xi^M_{\ell_1\ell_1'}(x_i,z) \text{cov}^{-1}_{\ell_1\ell_1'\ell_2\ell_2'}(x_i,x_j)  \tilde \xi^{M*}_{\ell_1\ell_1'}(x_j,z)\,,\nn
\end{align}
with 
\be
\frac{\partial}{\partial \alpha_i}\left(\sum_I \lambda_I Y_{2 M}(\hat{\omega}_I)\right)=\sum_I \lambda_I \left(\frac{\partial \theta_I}{\partial\alpha_i}\frac{\partial}{\partial\theta_I}+\frac{\partial \phi_I}{\partial\alpha_i}\frac{\partial}{\partial\phi_I}\right) Y_{2 M}(\theta_I, \phi_I)\,,
\ee
and $(\theta_I, \phi_I)$ are the polar angles defining the directions of $\hat{\omega}_I$.  We recall that
\begin{align}
\partial_{\theta} Y_{2 M}(\theta, \phi)&=-\frac{(\ipartial+\ipartial^*)}{2}Y_{2M}(\theta, \phi)=-\frac{\sqrt{6}}{2}\left(\,_1\!Y_{2M}(\theta, \phi)-\,_{-1}\!Y_{2M}(\theta, \phi)\right)\,,\nn\\
\partial_{\phi} Y_{2 M}(\theta, \phi)&=i \sin\theta\frac{(\ipartial-\ipartial^*)}{2}Y_{2M}(\theta, \phi)=i\sin\theta\frac{\sqrt{6}}{2}\left(\,_1\!Y_{2M}(\theta, \phi)+\,_{-1}\!Y_{2M}(\theta, \phi)\right)\,,
\end{align}
For definiteness, let us consider the case $\lambda_A=\lambda_1$ and $\alpha_i=\alpha$ in Eq. (\ref{Fisher1}). One has 
\be\label{Fisher2}
F_{\lambda_1, \alpha}=i \frac{\sqrt{6}}{2}\sum_M \left(Y_{2M}^*(\hat{\omega}_1)-Y_{2M}^*(\hat{\omega}_3)\right)\sum_I\lambda_I\sin\theta_I\left(_1Y_{2M}(\hat{\omega}_I)+\,_{-1}\!Y_{2M}(\hat{\omega}_I)\right)[\dots]\,,
\ee
where the $[\dots]$ represents the part of the Fisher matrix (\ref{Fisher1}) which does not depend on $\hat{\omega}_I$. 
We recall
\be\label{relation}
\sqrt{\frac{4\pi}{2\ell+1}}\sum_{m'} \,_m\!Y_{\ell m'}(\theta_1, \phi_1)\, _sY^*_{\ell m'}(\theta_2, \phi_2)= \,_sY^*_{\ell -m}(\beta, \alpha)e^{is\gamma}\,,
\ee
where here $(\alpha, \beta, \gamma)$ are the Euler angles of the rotation rotating the direction $(\theta_2, \phi_2)$ in $(\theta_1, \phi_1)$ and not the angles defined in Eq.~(\ref{eulwang}). In Eq. (\ref{Fisher2}) the products is between two harmonics evaluated either at the same directions or at orthogonal directions. In our case we have $\ell=2$, $s=0$ and $m=1$. Furthermore $(\beta,\alpha,\gamma)$ denotes a rotation by either $0$ or $\pi/2$ since either $\hat\om_1=\hat\om_2$ or they enclose and angle of $\pi/2$. In other words, $R(\beta,\alpha,\gamma)\eb_3= \pm\eb_I$ where $I\in\{1,2,3\}$ and $Y_{\ell m}(\beta, \alpha)= Y_{\ell m}(R^{-1}(\beta,\alpha,\gamma)\eb_3) = Y_{\ell m}(\pm\eb_I)$, see~\cite{RuthBook}.  The Euler angle $\gamma$ is irrelevant here since a rotation around $\eb_z$ leaves $\eb_z$ invariant.  But for the cartesian axes  $\eb_I$, $\vartheta$ is either $0$ or $\pi/2$ and $Y_{21}(\vartheta,\varphi)\propto \sin\vartheta\cos\vartheta$ vanishes. 
This completes the proof that the off-diagonal boxes in the Fisher matrix vanish.

\section{$\xi_{\ell\ell'}^{2M}$}\label{explapp}
The explicit expressions for the real-space version of Eqs.~(\ref{coeffa})-(\ref{coeffc}) are given by
\be
\xi_{\ell\ell'}^{2M\text{(a)}}  = -\frac{16 \pi^{3/2}}{45} C_{\ell}^{\Omega}(z,x)\sum\limits_I \lambda_I Y^*_{2M}(\hat \omega_I)  \left( \delta_{\ell,0}\delta_{\ell',2} +2\sqrt{\frac{2\ell'+1}{5}} \tj{2}{2}{\ell'}{0}{0}{0}\delta_{\ell,2} \right)   \,,
\label{realcoeffa}
\ee
\be
\begin{split}
\xi_{\ell\ell'}^{2M\text{(b)}}  =& -\frac{16 \pi^{3/2}}{5} C_{\ell}^{\Omega}(z,x)\sqrt{(2\ell+1)(2\ell'+1)}\sqrt{\frac{2}{15}}\sum\limits_I \lambda_IY^*_{2M}(\hat \omega_I) \\
&\times \bigg[ 2 \tj{3}{1}{\ell}{0}{0}{0}  \tj{3}{1}{\ell'}{0}{0}{0} \ninej{1}{2}{1}{\ell}{3}{\ell'} + 3 \tj{1}{1}{\ell}{0}{0}{0}  \tj{1}{1}{\ell'}{0}{0}{0} \ninej{1}{2}{1}{\ell}{1}{\ell'} \bigg]
\,,
\end{split}
\label{realcoeffb}
\ee
\be
\begin{split}
\xi_{\ell\ell'}^{2M\text{(c)}}  =&  -\frac{16 \pi^{3/2}}{15} C_{\ell}^{\Omega}(z,x)\sum\limits_I \lambda_I Y^*_{2M}(\hat \omega_I) \bigg[ \frac{1}{5} \delta_{\ell,2}\delta_{\ell',0}+ \frac{8}{105} \sqrt{2\ell+1}  \tj{4}{2}{\ell}{0}{0}{0} \delta_{\ell',4} \\
& + \frac{4}{7 \sqrt{5}}  \sqrt{2\ell+1}  \tj{2}{2}{\ell}{0}{0}{0} \delta_{\ell',2} \bigg]  \,.
\end{split}
\label{realcoeffc}
\ee
\end{widetext}

\bibliographystyle{utcaps}
\bibliography{VectorRefs}

\end{document}